\documentclass[english,aps,11pt,prd,superscriptaddress,preprintnumbers,nofootinbib]{revtex4}
\usepackage[T1]{fontenc}
\usepackage[latin9]{inputenc}
\usepackage{verbatim}
\usepackage{amsmath}
\usepackage{dcolumn}
\usepackage{graphicx}
\newcommand{\be}{\begin{equation}}
\newcommand{\ee}{\end{equation}}
\newcommand{\bea}{\begin{eqnarray}}
\newcommand{\eea}{\end{eqnarray}}
\newcommand{\ep}{\epsilon}

\usepackage{babel}
\begin{document}
\global\long\def\order#1{\mathcal{O}\left(#1\right)}
\global\long\def\d{\mathrm{d}}
\global\long\def\P{P}
\global\long\def\amp{{\mathcal M}}
\preprint{TTP15-013, CERN-PH-TH-2015-069}

\title{Two-loop helicity amplitudes  for the production of two off-shell 
electroweak bosons   in gluon fusion}

\author{Fabrizio Caola} 
\email{fabrizio.caola@cern.ch}
\affiliation{PH Department, TH Unit, CERN, 1211 Geneva 23, Switzerland}
\author{Johannes M. Henn}
\email{jmhenn@ias.edu}
\affiliation{Institute for Advanced Study, Princeton, NJ 08540, USA}
\author{Kirill Melnikov}
\email{kirill.melnikov@kit.edu}
\affiliation{ Institute for Theoretical Particle Physics, Karlsruhe Institute of Technology, 
Karlsruhe, Germany}
\author{Alexander V. Smirnov}
\email{asmirnov80@gmail.com}
\affiliation{
Research Computing Center, Moscow State University, 119991,
Moscow, Russia}
\author{Vladimir A. Smirnov}
\email{smirnov@theory.sinp.msu.ru}
\affiliation{
Skobeltsyn Institute of Nuclear Physics,  Moscow State University, 
119991 Moscow, Russia}

\begin{abstract}
\hspace*{-0.3cm}We compute the  part of the two-loop virtual amplitude for the process 
$gg  \to V_1 V_2  \to (l_1 \bar l'_{1})  (l_2 \bar l'_2)$, where $V_{1,2}$ are 
arbitrary electroweak gauge bosons, that receives contributions from loops of  massless quarks.  
Invariant masses of electroweak bosons are allowed to be  different from each other. 
Our result provides an important ingredient  for improving the
description of gluon fusion contribution to the production 
of four-lepton final states at the LHC.
\end{abstract}

\maketitle

\newpage 

\section{Introduction} 
Production of vector boson pairs   at the LHC is an important process for a variety of 
reasons. They include interesting checks of the Standard Model, 
searches for anomalous vector boson couplings and studies of a dominant background to Higgs 
boson production.  Since the vector boson pair production cross-section is currently  known 
to next-to-next-to-leading order (NNLO) in perturbative QCD~\cite{Cascioli:2014yka,tg1}, it is tempting to conclude 
that theoretical understanding of this process is already good enough and further improvements 
are unnecessary.  Unfortunately, such a conclusion would be premature. 

Indeed, since production of vector boson pairs  occurs predominantly through the annihilation 
of quark-antiquark pairs,  the one-loop gluon fusion matrix element contributes 
to $pp \to V_1V_2$ cross-section at NNLO {\it for the first time}.  The 
increase in the $pp \to V_1V_2$ total cross-section caused by the gluon fusion contribution 
is not very  large -- it is about five percent in case of the $W^+ W^-$ final state~\cite{tg1}. However, 
this gluon fusion contribution 
is somewhat larger than    the theoretical uncertainty in the production 
cross section, as estimated in recent 
NNLO QCD computations~\cite{Cascioli:2014yka,tg1}.
Therefore, for  these  uncertainty estimates to be valid,   the  current calculations of the 
gluon fusion contribution to $pp \to V_1 V_2$ must be  accurate to $30-50$ percent.   
As we explain in the next paragraph, it is conceivable  that this is  not the case.
In addition,  the relevance of the gluon-fusion contribution  strongly depends on the applied cuts and selection criteria. 
For example,  typical cuts used by ATLAS and CMS in Higgs bosons searches increase the fraction of $gg\to V_1 V_2$ 
contribution to the background cross-section to ${\cal O}(10\%)$ \cite{Cascioli:2013gfa}. 
With more aggressive cuts, this fraction can increase to 
an astounding $30\%$, as shown for instance in Ref.~\cite{Binoth:2006mf}.
Good theoretical control of the gluon fusion process becomes very  important in this case.

Since the gluon fusion contribution may play an important role in four-lepton production at hadron colliders, 
it is important to realize that  radiative corrections to $gg \to V_1 V_2$   can {\it naturally} be  large. Indeed,  
$gg \to V_1 V_2$ is  a  process where 
an initial  state with large color charge annihilates  into a colorless final state.  NLO  QCD radiative corrections 
computed for similar processes, such as $gg \to H$, $gg \to HH$, $gg \to HZ$  
and $gg \to \gamma \gamma$~\cite{Spira:1995rr,Dawson:1998py, Altenkamp:2012sx,Bern:2002jx}
 turned out to be as large as  $50-100$ percent.   Therefore, to fully  control the gluon 
fusion contribution to vector boson pair production, NLO QCD corrections to $gg \to V_1 V_2$ need to be calculated. 
The first step in this direction is the calculation  of the two-loop (next-to-leading order) scattering amplitude  
$gg \to V_1 V_2$.  Once the two-loop scattering amplitude becomes  available, it will be possible to compute 
the  gluon fusion contribution to vector boson pair production cross-section  through NLO in perturbative QCD by supplementing 
it with the real emission contribution $gg \to V_1 V_2 + g$. 

The goal  of the present  paper is to present  a  calculation of the $gg \to V_1 V_2$ helicity amplitudes {\it mediated 
by loops of massless quarks.}   We choose to work within the  massless approximation  because 
in this case all the relevant two-loop integrals are known~\cite{planar,nonplanar,Papadopoulos:2014hla,lorenzo}\footnote{
For earlier results in the case of equal-mass vector bosons, see~\cite{Gehrmann:2014bfa,Gehrmann:2013cxs}.}
and because in many cases massless quarks provide 
significant fraction of the full amplitude. 
We note, however, that there   are phenomenological situations where 
top quark loops are important contributors to $gg \to V_1 V_2$~\cite{Campbell:2011cu}; our calculation is not 
applicable to those cases since for top quarks the
massless approximation is obviously invalid.  We note in this regard that a proper treatment of 
massive quarks  in loop-mediated  processes is an important and difficult problem whose solution 
is not known at the moment.  We therefore start with the massless quark contribution 
to the gluon fusion process $gg \to V_1 V_2$ which, at the very least, will allow an informed  estimate of the magnitude 
of NLO QCD corrections. Our result can then be supplemented with the contribution of the massive top quark loop once 
it becomes available.  First steps in that direction where recently reported in 
Refs.~\cite{Melnikov:2015laa}.

The remainder of the paper is organized as follows. 
In Section~\ref{section1} we discuss the set-up of the calculation and the parametrization of the $gg \to V_1 V_2$ 
amplitude.  In Section~\ref{section2}  we describe numerical implementation and checks 
of the calculated amplitudes and present some numerical results.  
We conclude in Section~\ref{concl}.  Finally, we note that  
 many aspects of the calculation that we report here are similar to what we  described in 
Ref.~\cite{Caola:2014iua}; nevertheless we believe that there are essential  differences in the calculation 
of $q \bar q \to V_1 V_2$ \cite{Caola:2014iua,lorenzo} and $g g \to V_1 V_2$ amplitudes at two loops 
to warrant a discussion of the latter in a separate 
publication.

\section{The set up of the computation} 
\label{section1}

We consider the process 
$g(p_1) g(p_2) \to ( V_1^*(p_3) \to l(p_5) \bar l(p_6) ) 
( V_2^*(p_4) \to l(p_7) \bar l(p_8) ) 
$.
We work in the approximation where quarks of the first two generations are massless and quarks of the 
third generation are neglected. The CKM matrix is taken to be an identity matrix.  We write the matrix element 
as 
\be
{\cal M}(\lambda_{g_1} , \lambda_{g_2}, \lambda_5, \lambda_7) = i\left ( \frac{g_W}{\sqrt{2}} \right )^4 
  \delta^{a_1 a_2} {\cal D}_3 {\cal D}_4  C_{l,V_2}^{\lambda_7} C_{l,V_1}^{\lambda_5} 
\epsilon_3^\mu(\lambda_5)  \epsilon_4^\nu(\lambda_7) \;
  C_{V_1 V_2} \;  {\cal A}_{\mu \nu}(p_1^{\lambda_{g_1}},p^{\lambda_{g_2}}_2; p_3, p_4), 
\label{eq21}
\ee
where $a_{1,2}$ are the color indices of the incoming gluons, $g_W = e/\sin \theta_W$ is the $SU(2)$ weak coupling,  
${\cal D}_i = 1/(p_i^2 - m_{V_i}^2 + i m_{V_i} \Gamma_{V_i})$ is the $V_i$-boson propagator,
$\lambda_{g_1,g_2}, \lambda_5, \lambda_7$ are helicities  of the incoming gluons and outgoing leptons, 
  $C_{l,V_2}^{\lambda_7} C_{l,V_1}^{\lambda_5} $ are 
helicity-dependent 
couplings of vector bosons to  leptons,
 and  $\epsilon_{3,4}$  are matrix elements for leptonic decays of $V_1$ and $V_2$ that we will 
specify shortly.   The factor $C_{V_1 V_2}$ involves sums of couplings of virtual fermions 
to gauge bosons.    

We now give   these couplings explicitly. The couplings of electroweak vector bosons to external  leptons  read
\be
C_\gamma^{L,R} = -\sqrt{2} Q_l \sin \theta_W,
\;\;\;\;\;C_{l,Z}^{L,R} = \frac{1}{\sqrt{2} \cos \theta_W} \left ( V_l \pm A_l \right ), 
\;\;\;\;\; C_{lW^+}^\lambda = C_{l W^-}^\lambda = \delta_{\lambda L}.
\ee
Here, $V_e = -1/2+2 \sin^2 \theta_W$, $V_\nu = 1/2$, $A_e = -1/2$,  $A_\nu = 1/2$ and 
$Q_l$ is the lepton electric charge in units of the positron charge.  
The couplings to virtual quarks, where two massless generations are included,  are 
\be
\begin{split} 
& C_{\gamma \gamma} = \frac{20 \sin \theta_W^2}{9},
\;\;\;\;\;\;\;\;
C_{ZZ} = \frac{\left ( V_u^2 + V_d^2 + A_u^2 + A_d^2 \right )}{\cos \theta_W^2}, 
\;\;\;
\\
& C_{Z \gamma} = -\frac{2 \sin \theta_W}{\cos \theta_W} \left ( V_u Q_u + V_d Q_d \right ),
\;\;\;\;\;\;\;\;
C_{W^+W^-} = 1, 
\end{split} 
\label{eq41}
\ee
where 
$Q_{u,d}$ are the electric charges of up and down quarks  in units of the positron charge and 
$ V_u = 1/2 - 4/3 \sin^2 \theta_W$, $A_u = 1/2$, $V_d = -1/2 + 2/3 \sin^2 \theta_W$, $A_d = -1/2$. 

 A word of caution is required here. As can be seen from Eq.~(\ref{eq41}), the amplitude for the gluon fusion 
into a pair of vector bosons is proportional to sums of squares of vector and axial-vector couplings of these vector 
bosons to internal fermions. This means that  diagrams with two vector currents and  diagrams 
with two axial  currents give identical contributions to the amplitude 
and that  diagrams that involve one vector current  and one axial current 
do not contribute at all.  These two  features have different origins. The first 
one is  the result of the massless approximation and an ensuing relations between matrix elements 
of vector and axial currents.  The second feature is a direct consequence of  $C$-parity conservation~\cite{ax}, that enforces the cancellation 
of these  axial-vector  terms when contributions  of all diagrams are summed up. 
Therefore, the  knowledge of the amplitude for  vector-like couplings 
of gauge bosons to quarks is sufficient to reconstruct 
the gluon fusion  amplitude for  electroweak gauge bosons whose couplings to fermions are linear 
combinations of vector and axial-vector couplings. 

The  scattering amplitude in Eq.~(\ref{eq21}) stripped of all non-essential couplings reads 
\be
{\cal A} = {\cal A}_{\mu \nu}  \epsilon_3^\mu  \epsilon_4^\nu  
= {\bar A}_{\mu \nu \alpha \beta} \epsilon_{1,\lambda_1}^\mu \epsilon_{2,\lambda_2}^\nu \epsilon_3^\alpha \epsilon_4^\beta,
\ee
where $\epsilon_{1,2}$ are polarization vectors for gluons with definite  helicities. As we just explained, 
we need to compute the amplitude ${\cal A}$ for vector-like interactions of massive vector bosons. 
We need to decompose the amplitude ${\cal A}$ into invariant form factors that are independent of 
polarization vectors.  To do this, we choose physical polarizations for gluons and vector bosons, 
\be
\epsilon_1 \cdot p_{1,2} = 0, \;\;\;\; \epsilon_{2} \cdot p_{1,2} = 0, 
\;\;\;\;
\epsilon_3 \cdot p_3 = 0, 
\;\;\;\;\;
\epsilon_4 \cdot p_4 = 0.
\label{eq2}
\ee
It follows from Eq.(\ref{eq2})  that we use $p_{1,2}$ as a reference vector for polarization vectors $\epsilon_{2,1}$, 
respectively. 

To reduce the number of independent scalar 
products, we employ the Sudakov decomposition of the vector boson momenta,  
$p_{3,4} = \alpha_{3,4} p_1 + \beta_{3,4} p_2 \pm p_\perp$. The coefficients 
$\alpha_{3,4}$ and $\beta_{3,4}$ are expressed in terms of Mandelstam invariants 
\be
\alpha_3 = \frac{m_3^2 - u}{s},\;\;\; 
\beta_3 = \frac{m_3^2 - t}{s},
\;\;\;
\alpha_4 = \frac{m_4^2 - t}{s},
\;\;\;
\beta_4 = \frac{m_4^2 - u}{s}.
\label{eq61}
\ee

 Since  $\epsilon_3 \cdot  p_3 = 0$ and $\epsilon_4 \cdot p_4 = 0$,   
we can choose $\epsilon_{3,4} \cdot p_1$ and $\epsilon_{3,4} \cdot p_2$ as independent scalar products; 
$\epsilon_{3,4} \cdot p_\perp$ are then given by linear combinations of $\epsilon_{3,4} \cdot p_{1,2}$.
Given these constraints,  the scattering amplitude is represented by  the following expression 
\be
\label{eq4}
\begin{split}
& {\cal A} = 
T_1 \left ( \epsilon_1 \cdot \epsilon_2 \right ) \left ( \epsilon_3 \cdot \epsilon_4  \right ) 
+
T_2 \left ( \epsilon_1 \cdot \epsilon_3 \right ) \left ( \epsilon_2 \cdot \epsilon_4  \right ) 
+
T_3 \left ( \epsilon_1 \cdot \epsilon_4 \right ) \left ( \epsilon_2 \cdot \epsilon_3  \right ) 
\\
& + \left ( \epsilon_1 \cdot \epsilon_2 \right ) \left ( 
T_4  ( p_1 \cdot \epsilon_3) \left ( p_1 \cdot \epsilon_4 \right ) 
+
T_5  ( p_1 \cdot \epsilon_3) \left ( p_2 \cdot \epsilon_4 \right ) 
+
T_6  ( p_2 \cdot \epsilon_3) \left ( p_1 \cdot \epsilon_4 \right ) 
+
T_7  ( p_2 \cdot \epsilon_3) \left ( p_2 \cdot \epsilon_4 \right ) 
\right )
\\
& +
\left ( \epsilon_1 \cdot \epsilon_3 \right ) \left ( p_\perp \cdot \epsilon_2 \right ) 
\left ( T_8( p_1 \cdot \epsilon_4) + T_9 \left ( p_2 \cdot \epsilon_4 \right )  \right ) 
+
\left ( \epsilon_1 \cdot \epsilon_4 \right ) \left ( p_\perp \cdot \epsilon_2 \right ) 
\left ( T_{10}( p_1 \cdot \epsilon_3) + T_{11} \left ( p_2 \cdot \epsilon_3 \right )  \right ) 
\\
& +\left ( \epsilon_2 \cdot \epsilon_3 \right ) \left ( p_\perp \cdot \epsilon_1 \right ) 
\left ( T_{12}( p_1 \cdot \epsilon_4) + T_{13} \left ( p_2 \cdot \epsilon_4  \right )  \right ) 
+
\left ( \epsilon_2 \cdot \epsilon_4 \right ) \left ( p_\perp \cdot \epsilon_1 \right ) 
\left ( T_{14}( p_1 \cdot \epsilon_3 ) + T_{15} \left ( p_2 \cdot \epsilon_3 \right )  \right ) 
\\
&
+ \left ( \epsilon_1 \cdot p_\perp \right ) \left ( \epsilon_2 \cdot p_\perp \right )  
\left ( 
T_{17} ( p_1 \cdot \epsilon_3) \left ( p_1 \cdot \epsilon_4 \right ) 
+
T_{18}  ( p_1 \cdot \epsilon_3) \left ( p_2 \cdot \epsilon_4 \right ) 
+
T_{19}  ( p_2 \cdot \epsilon_3) \left ( p_1 \cdot \epsilon_4  \right ) 
\right.\\
&
\left.
+ T_{20}  ( p_2 \cdot \epsilon_3) \left ( p_2 \cdot \epsilon_4 \right ) 
\right )
+
 \left ( \epsilon_3 \cdot \epsilon_4 \right ) \left ( p_\perp \cdot \epsilon_1 \right ) \left ( p_\perp \cdot \epsilon_2 \right ) T_{16} 
.
\end{split}
\ee
The form-factors $T_{1..20}$ are functions of Mandelstam kinematic variables $s,t,u$ and  invariant masses of the two 
vector bosons $m_{3,4}$. 

The problem with the calculation of the scattering amplitude ``as it is'' is that carrying around polarization vectors for external gluons 
and vector bosons is extremely  expensive for computations that employ  integration-by-parts identities~\cite{ibp,ibp1}, 
 the main  vehicle  for multi-loop calculations.  For this reason, we need a procedure  that allows us to compute 
amplitudes  but where no  vectors,  except for momenta of external particles, appear in the calculation.   
A suitable method   employs projection operators.  The idea is to  compute 
the amplitude ${\cal A}$ in Eq.~(\ref{eq4}) with polarization vectors $\epsilon_{1..4}$ substituted by linear combinations 
of external momenta $p_{1,2,3,4}$.  This procedure allows us to compute the ``projected'' amplitude ${\cal A}$ in terms of
form factors $T_{1...20}$ and  also using integration-by-parts identities since polarization vectors 
completely disappear  from the calculation.  By choosing a sufficient number of independent ``projection operators'', 
we produce a system of equations that we can solve for the form factors $T_{1..20}$. 

However, there is a subtlety.  
The expression for the amplitude in Eq.~(\ref{eq4}) was written under the assumption that polarization vectors 
satisfy the transversality conditions Eq.~(\ref{eq2}); those conditions will be definitely violated if polarization vectors are replaced 
by linear combinations of external momenta.  To get around this problem, we write the amplitude as
\be
{\cal A}  ={\bar A}_{\mu \nu \alpha \beta} 
{\cal P}_{12}^{\mu \mu_1} {\cal P}_{12}^{\nu \nu_1} {\cal P}_{3}^{\alpha \alpha_1} {\cal P}_{4}^{\beta \beta_1}
\epsilon_{1 \mu_1}  \epsilon_{2 \nu_1} \epsilon_{3 \alpha_1} \epsilon_{4 \beta_1},
\label{eq5}
\ee
where
\be
{\cal P}_{12}^{\mu \nu} = -g^{\mu \nu} + \frac{p_1^\mu p_2^\nu + p_1^\nu p_2^\mu}{p_1 \cdot p_2},
\;\;\;
{\cal P}_3^{\mu \nu}  = -g^{\mu \nu} + \frac{p_3^\mu p_3^\nu}{p_3^2},
\;\;\;
{\cal P}_4^{\mu \nu}  = -g^{\mu \nu} + \frac{p_4^\mu p_4^\nu}{p_4^2}.
\ee
We then note that in Eq.~(\ref{eq5}) we can use {\it any} vector $\epsilon_i$ to calculate the amplitude since 
projection  operators ${\cal P}_{12}$, ${\cal P}_{3,4}$ automatically ensure the transversality constraints $p_{1,2} \cdot \epsilon_1 = 0$, 
$p_{1,2} \cdot \epsilon_2 = 0$,  $p_3 \cdot \epsilon_3 = 0$ and $p_4 \cdot \epsilon_4 = 0$.  We can also replace 
$\epsilon_1^\mu \epsilon_2^\nu \epsilon_3^\alpha \epsilon_4^\beta$ with an arbitrary rank-four tensor since 
it will also be projected on the appropriate  transverse space.

We now list all  projection  operators that we use in the computation of $gg \to V_1 V_2 $ amplitude. 
To this end, we define 
the  amplitude contracted with  the projection operators 
\be
{\cal O}^{\mu_1 \mu_2 \mu_3 \mu_4} = {\bar A}_{\nu_1 \nu_2 \nu_3 \nu_4} 
{\cal P}_{12}^{\nu_1 \mu_1} {\cal P}_{12}^{\nu_2 \mu_2} {\cal P}_{3}^{\nu_3 \mu_3} {\cal P}_{4}^{\nu_4 \mu_4}.
\ee
We also define a tensor that is a projector on the vector space that is orthogonal to the collision plane. It reads 
\be
t^{\mu}_{ \nu} =   \delta_{\nu  p_1 p_2 p_\perp}^{\mu p_1 p_2 p_\perp},
\ee
where $ \delta_{\nu_1 p_1 p_2 p_\perp }^{\mu_1 p_1 p_2 p_\perp} = \delta_{\nu_1 \nu_2 \nu_3 \nu_4}^{\mu_1 \mu_2 \mu_3 \mu_4 } \;  p_1^{\nu_2}  p_2^{\nu_3} p_\perp^{\nu_4}  
p_{1, \mu_2}  p_{2, \mu_3} p_{\perp ,\mu_4}$ and 
\be
\label{eq_det}
\delta_{\nu_1 \nu_2 \nu_3 \nu_4}^{\mu_1 \mu_2 \mu_3 \mu_4 }
 = {\rm det} | g^{\mu_{i \in \{1...4\}}}_{\nu_{j \in \{1..4\}}} |.
\ee  
It is clear that contraction of tensor $t_{\mu \nu}$ with {\it any}  linear combination 
of $p_1, p_2$ and $p_\perp$ vanishes thanks to the antisymmetry 
of the determinant in Eq.~(\ref{eq_det}).   

We define  twenty projections of the amplitude ${\cal A}$ on linear combinations of $T_{1...20}$ 
by making different choices of  the ``polarization vectors''. They are  
\be
\begin{split}
& G_1 = {\cal O}^{\mu_1\mu_2\mu_3\mu_4} g_{\mu_1\mu_2} g_{\mu_3\mu_4},\;\;\;\;\;\;\;\;\;\;\;\;\;\;\;\;
G_2 = {\cal O}^{\mu_1\mu_2\mu_3\mu_4} g_{\mu_1\mu_3} g_{\mu_2\mu_4} , \;\;\;\;\;\;\
\\
& G_3 = {\cal O}^{\mu_1\mu_2\mu_3\mu_4} g_{\mu_1\mu_4} g_{\mu_2\mu_3},\;\;\;\;\;\;\;\;\;\;\;\;\;\;\;\;
G_4 = p_\perp^{-4} s^{-2}{\cal O}^{p_\perp p_\perp p_1 p_1 },
\\
&  G_5 =  p_\perp^{-4} s^{-2} {\cal O}^{p_\perp p_\perp p_1 p_2},\;\;\;\;\;\;\;\;\;\;\;\;\;\;\;\;\;\;\;\;\;
 G_6 = p_\perp^{-4} s^{-2} {\cal O}^{p_\perp p_\perp p_2 p_1},\;\;\;\;\;
\\
&
G_7 =  p_\perp^{-4} s^{-2} {\cal O}^{p_\perp p_\perp p_2 p_2  },\;\;\;\;\;\;\;\;\;\;\;\;\;\;\;\;\;\;\;\;\;
G_8 = 4 p_\perp^{-6} s^{-2} {\cal O}^{p_\perp p_\perp   \mu_3  \mu_4 }   t_{\mu_3 \mu_4},
\\
& G_9 =4 p_\perp^{-6} s^{-6} {\cal O}^{\mu_1 \mu_2  p_1 p_1  }  t_{\mu_1 \mu_2} ,\;\;\;\;\;\;\;\;\;\;\;\;
G_{10} = 8 p_\perp^{-4} s^{-3} {\cal O}^{p_\perp \mu_2  \mu_3  p_1}   t_{\mu_2 \mu_3},
\\
&  G_{11} = 4 p_\perp^{-6} s^{-3} {\cal O}^{p_\perp \mu_2  \mu_3  p_\perp  }  t_{\mu_2 \mu_3} ,\;\;\;\;\;\;\;\;\;\;
 G_{12} = 8 p_\perp^{-4} s^{-3} {\cal O}^{\mu_1 p_\perp  p_1  \mu_4 }  t_{\mu_1, \mu_4}, 
\\
&
 G_{13} = 4p_\perp^{-6}s^{-3}  {\cal O}^{\mu_1 p_\perp  p_\perp   \mu_4 }  t_{\mu_1, \mu_4},\;\;\;\;\;\;\;\;\;
 G_{14} = 8 p_\perp^{-4} s^{-3}  {\cal O}^{\mu_1 p_\perp  \mu_3  p_2 }   t_{\mu_1, \mu_3},
\\
&  G_{15} =  4p_\perp^{-6}s^{-3} {\cal O}^{\mu_1 p_\perp  \mu_3  p_\perp}   t_{\mu_1, \mu_3}, \;\;\;\;\;\;\;\;\;
G_{16} = 8 p_\perp^{-4} s^{-3}   {\cal O}^{p_\perp \mu_2  p_1  \mu_4 }   t_{\mu_2, \mu_4},
\\
&   G_{17} = 4p_\perp^{-6}s^{-3} {\cal O}^{p_\perp \mu_2  p_\perp  \mu_4 }   t_{\mu_2, \mu_4},\;\;\;\;\;\;\;\;\;
  G_{18} = 4 p_\perp^{-6} s^{-6}  {\cal O}^{\mu_1 \mu_2  p_1  p_2 }  t_{\mu_1 \mu_2} ,
\\
&
G_{19} =4 p_\perp^{-6} s^{-6}   {\cal O}^{\mu_1 \mu_2  p_2   p_1 }  t_{\mu_1 \mu_2},\;\;\;\;\;\;\;\;\;\;\;
G_{20} = 4 p_\perp^{-6} s^{-6}  {\cal O}^{\mu_1 \mu_2  p_2  p_2 }  t_{\mu_1 \mu_2}.
\label{eqproj}
\end{split} 
\ee
In these equations, we used a simplified notation for the contraction of the tensor 
${\cal O}$ with a vector $a$, ${\cal O}^{\mu_1..\mu...\mu_n} a_\mu = {\cal O}^{\mu_1 ..a .. \mu_n}$.
Since $G_{1..20}$ only depend on scalar products of external momenta  and on scalar products 
of external momenta and the loop momenta, we can express    $G_{1..20}$ through known master 
integrals~\cite{nonplanar,planar} 
by applying  integration-by-parts identities~\cite{ibp,ibp1}. At the same time, 
the form factors $T_{1..20}$ can be written as linear combinations 
of $G_{1..20}$ in a straightforward way. 

To determine physical amplitude for the process $g(p_1) g(p_2) \to ( V^*(p_3) \to l(p_5) \bar l(p_6) ) 
( V^*(p_4) \to l(p_7) \bar l(p_8) ) $ we use spinor-helicity notations. 
Specifically, for the incoming gluons we choose
\be
\begin{split}
& \epsilon_{1L}^\mu =  -\frac{[ 2  \gamma^\mu  1 \rangle}{ \sqrt{2} [21]},
\;\;\; \epsilon_{1R}^\mu = \frac{ \langle 2 \gamma^\mu 1 ]}{\sqrt{2} \langle 21 \rangle },
\;\;\;\;
\epsilon_{2L}^\mu =  -\frac{[ 1  \gamma^\mu  2 \rangle}{ \sqrt{2} [12]},
\;\;\; \epsilon_{2R}^\mu = \frac{ \langle 1 \gamma^\mu 2 ]}{\sqrt{2} \langle 12 \rangle }.
\end{split} 
\ee

Since a complex conjugation of the helicity amplitudes reverses all helicities, we only 
need to consider two, rather than four, cases of equal and unequal helicities. We choose 
$L_1L_2$ and  $L_1 R_2$ as polarization states for gluons $g_1$  and $g_2$, respectively. 
We leave the polarization vectors of the massive vector bosons unspecified at this point. 
To proceed further, it is convenient to  write tensor products of gluon polarization vectors 
as follows  ( see e.g. Ref.~\cite{Binoth:2006mf})
\be
\label{eq_tensor}
\begin{split} 
& \ep_{1L,\mu } \ep_{2L,\nu} = \frac{\langle 1 2 \rangle  }{[1 2] s }
\left ( p_{1,\mu}p_{2,\nu} + p_{1,\nu} p_{2, \mu} 
 - g_{\mu, \nu} p_1 \cdot p_2 + i \epsilon_{\mu \nu p_1 p_2 } 
\right ),
\\
& \ep_{1L,\mu} \ep_{2R,\nu} = \frac{\langle 1 p_\perp  2 ]  }{[1 p_\perp 2 \rangle  p_\perp^2  s^2}
\left (
\frac{p_{\perp, \mu} p_{\perp \nu} s^2}{4} 
+ s i \epsilon_{p_1 p_2 p_\perp  \mu}  p_{\perp, \nu}
           - \epsilon_{p_1 p_2 p_\perp  \mu} \epsilon_{p_1 p_2 p_\perp  \nu} 
+ ( \mu \leftrightarrow \nu )
\right ).
\end{split} 
\ee
The transverse momentum $p_\perp^\mu$ is introduced  just before  Eq.~(\ref{eq61}). 
We use Eq.~(\ref{eq_tensor}) to express 
the amplitude ${\cal A}$ in Eq.~(\ref{eq4}) through 
nine independent Lorentz structures 
\be
\begin{split} 
{\cal A}_{\lambda_1 \lambda_2 } & =  {\cal N}_{\lambda_1 \lambda_2} \Bigg [  F^{\lambda_1 \lambda_2}_1  
\left( p_1 \cdot \ep_4 \right) \left( p_1 \cdot \ep_ 3 \right)
       + F^{\lambda_1 \lambda_2}_2  \left(p_1 \cdot \ep_4\right)\left( p_2 \cdot \ep_3 \right)
       + F^{\lambda_1 \lambda_2}_3 \left(p_1 \cdot \ep_3\right)\left( p_2 \cdot \ep_4 \right)
\\
& 
+ F^{\lambda_1 \lambda_2}_4  \left(p_2 \cdot \ep_ 4 \right)\left(p_2 \cdot \ep_3 \right)       + F^{\lambda_1 \lambda_2}_5  \ep_4 \cdot \ep_3 
+ i   \epsilon_{\mu \nu \alpha \beta} p_1^\mu p_2^\nu p_\perp^\alpha \ep_4^\beta 
 \left ( F^{\lambda_1 \lambda_2}_6 p_1 \cdot \ep_3   + F^{\lambda_1 \lambda_2}_7 p_2 \cdot \ep_3   \right )
\\
& + i   \epsilon_{\mu \nu \alpha \beta} p_1^\mu p_2^\nu p_\perp^\alpha \ep_3^\beta 
 \left ( F^{\lambda_1 \lambda_2}_8 p_1 \cdot \ep_4   + F^{\lambda_1 \lambda_2}_9 p_2 \cdot \ep_4   \right ) \Bigg ].
\label{eqa}
\end{split}
\ee
In Eq.(\ref{eqa}) $N_{\lambda_1 \lambda_2} $ are the normalization factors for left-left and left-right polarization cases
\be
{\cal N}_{LL} = \frac{\langle 1 2 \rangle }{ [12] s},
\;\;\;\;\;\;\;\; 
{\cal N}_{LR} = \frac{\langle 1 \hat p_\perp 2 ] }{[1 \hat p_\perp 2 \rangle p_\perp^2  s^2},
\ee
and  $F^{\lambda_1 \lambda_2}_{i =1..9}$ are helicity-dependent form factors that are functions 
of the Mandelstam variables and the invariant masses of vector bosons. 

To account for transitions of vector bosons to final state leptons, 
their  polarization vectors are  replaced by  matrix elements of vector and axial-vector currents.
We therefore choose 
\be
\begin{split} 
\epsilon_{3L}^{\mu} = \langle 5 | \gamma^\mu | 6], \;\;\;
\epsilon_{3R}^{\mu} = \langle 6 | \gamma^\mu | 5], \;\;\;
\;\;\;\;\epsilon_{4L}^{\mu} = \langle 7 | \gamma^\mu | 8], \;\;\;
\epsilon_{4R}^{\mu} = \langle 8 | \gamma^\mu | 7].
\end{split} 
\ee

We note that, although we need helicity amplitudes for all possible helicity combinations 
of leptons,  it is sufficient to compute just one of them since other helicity amplitudes 
can be obtained using simple replacement rules. 
The amplitudes for left-handed polarization of both electroweak vector bosons read
\be
\begin{split} 
{\cal A}^{\lambda_1 \lambda_2}_{3L 4L} = & {\cal N}_{\lambda_1 \lambda_2}  \Big \{
  \Big ( F^{\lambda_1 \lambda_2}_1 \langle 1 5 \rangle    [6 1] 
             + F^{\lambda_1 \lambda_2}_2   \langle 2 5 \rangle  [6 2] \Big ) \langle 1 7 \rangle  [8 1] 
\\
&   
+ \Big ( F^{\lambda_1 \lambda_2}_3 \langle 1 5 \rangle  [6 1] 
+ F^{\lambda_1 \lambda_2}_4 \langle 2 5 \rangle  [6 2] \Big ) 
\langle 2 7 \rangle [8 2] 
   + 2 F^{\lambda_1 \lambda_2}_5 \langle 5 7 \rangle [8 6] 
\\
& 
+  \frac{1}{2} \Big ( F^{\lambda_1 \lambda_2}_6 \langle 1 5 \rangle  [6 1] 
+ F^{\lambda_1 \lambda_2}_7 \langle 2 5 \rangle [6 2] \Big )
  \Big (  \langle 1 2 \rangle \langle 7 8 \rangle [8 1] [8 2] +
     \langle 1 7 \rangle \langle 2 7 \rangle [2 1] [8 7]  \Big  )  
\\
&   -\frac{1}{2} 
\Big (F^{\lambda_1 \lambda_2}_8 \langle 1 7 \rangle [8 1]
+ F^{\lambda_1 \lambda_2}_9 \langle 2 7 \rangle [8 2]     
\Big )
\Big ( \langle 1 2 \rangle \langle 5 6 \rangle [6 1] [6 2]   
     +  \langle 1 5 \rangle \langle 2  5 \rangle [2 1] [6 5] \Big )
\Big \}. 
\end{split}
\label{amplhel}
\ee
Amplitudes for right-handed polarizations of the vector boson with momentum $p_3$ ($p_4$) are obtained from the above ones
upon the replacement $5\leftrightarrow 6$ ($7\leftrightarrow 8$). Finally, all remaining  helicity amplitudes 
can be obtain by replacing all angle brackets in spinor products 
with square brackets in Eq.(\ref{amplhel}) and vice versa
\be
{\cal A}^{-\lambda_1-\lambda_2}_{3R4R} = {\cal A}^{\lambda_1\lambda_2}_{3L4L}\left[{\langle ij \rangle \leftrightarrow [ij]}\right]
\ee

The $F$ form factors that enter the amplitudes are expressed through either $T$ or $G$ form factors.  
This can be done in a straightforward way using  Eqs.~(\ref{eq4},\ref{eqproj},\ref{eq_tensor},\ref{eqa}).  Examples of corresponding relations 
are given in the Appendix. 

\begin{figure}[t]
  \centering
  \includegraphics[angle=0,width=0.75\textwidth]{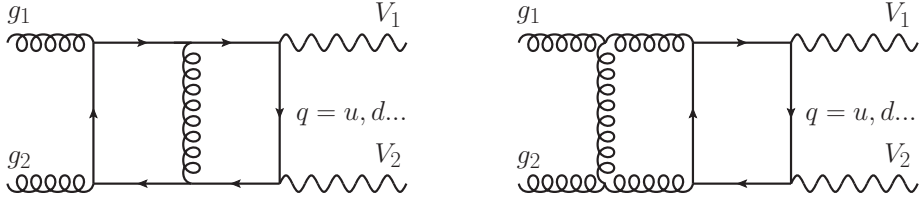}\\
  \caption{Representative two-loop diagrams that describe  production of vector boson pairs 
in gluon fusion. }
  \label{fig1}
\end{figure}

\section{Calculation of the amplitude} 
\label{section2}

We apply the set-up described in the  previous Section to   the calculation of  gluon-fusion amplitude.  
There are 93 non-vanishing two-loop diagrams that contribute to the $gg \to VV$ amplitude; some examples 
are shown in Fig.~\ref{fig1}. We generate 
the relevant diagrams using {\sf QGRAF}~\cite{qgraph}
 and process them with {\sf Maple} and  {\sf Form}~\cite{form4}. 
We compute the contribution of every diagram to the $G$ and eventually $F$ form factors.
At this point, the result is expressed in terms of two-loop tensor integrals. These integrals can be 
classified in terms of six different topologies, three of which are planar and three are non-planar~\cite{planar,nonplanar}.
The tensor integrals are expressed through the master integrals computed in Refs.~\cite{planar,nonplanar}, 
using integration-by-parts technology~\cite{ibp,ibp1}.  We employ the program {\sf FIRE}~\cite{Smirnov:2008iw,Smirnov:2013dia,Smirnov:fire}
to achieve this.  
Combining contributions of different diagrams, we obtain 
the results for the eighteen form factors (nine for $LL$ gluon helicity configuration and nine for $LR$ gluon 
helicity configuration) that are required to describe all helicity amplitudes for $gg \to V_1 V_2$ process. 
We note that, compared to the calculation of $q \bar q \to V_1 V_2$ amplitude, the case of $gg \to V_1 V_2$ 
requires more complicated  reduction since tensor integrals of a higher rank appear.  Nevertheless, {\sf FIRE} 
can successfully deal with this challenge. 

As we already mentioned, the helicity amplitudes are expressed in terms of master integrals computed in 
Refs.~\cite{planar,nonplanar}. The analytic expressions for these master integrals involve various 
functions, including logarithms, polylogarithms of multiple ranks as well as generalized Goncharov 
polylogarithms.  To compute the latter, we use their  numerical implementation~\cite{Vollinga:2004sn} 
in the computer algebra program {\sf GiNaC}~\cite{Bauer:2000cp}. We note that {\sf GiNaC} can be called from both 
{\sf Mathematica} and {\sf Fortran} providing  multiple options for the numerical evaluation of the amplitude. 

The $gg \to V_1 V_2$ amplitude appears for the first time at one loop; for this reason this amplitude is 
ultraviolet  and infra-red finite.  The two-loop $gg \to V_1V_2$ amplitude  contains at most ${\cal O}(1/\ep^2)$  singularities, 
where $\ep = (4-d)/2$ is the parameter of dimensional regularization.  The divergences of the two-loop 
$gg \to V_1V_2$ amplitude  can be predicted in terms of the one-loop amplitude using results of Ref.~\cite{Catani:1998bh}.
The  relation between  one- and two-loop amplitudes becomes very simple if expressed through bare, rather than renormalized QCD  
coupling.  It reads 
\be
{\cal A}_2 = -\frac{C_A}{\ep^2}  e^{\ep i \pi  } {\cal A}_1 + {\cal O}(\ep^0),
\label{eqdiv}
\ee
where $C_A = 3$ is the QCD color factor and ${\cal A}_{1,2}$ are defined through the following expression for the amplitude
\be
{\cal A} =  \frac{a_0}{2}  
\left [  s^{-\ep } \;{\cal A}_{1}  + a_0 s^{-2\ep} {\cal A}_2 + {\cal O}(\alpha_s^2) \right ].
\label{eq_ampl}
\ee
In Eq.~(\ref{eq_ampl}) we use 
$ a_0 = \alpha_s^{(0)} \Gamma(1+\ep) (4\pi)^\ep/(2\pi)$, where   $\alpha_s^{(0)}$ is the {\it bare} QCD coupling constant.

A connection between divergences of the two-loop amplitude ${\cal A}_2$ and the one-loop amplitude 
${\cal A}_1$ given by Eq.~(\ref{eqdiv}) 
is important for checks of the correctness of the calculation since the  computation of 
${\cal A}_2$ proceeds without  
separation into divergent and convergent parts until the very end.   

We are now in position to present some numerical results for the $gg \to V_1 V_2$ amplitude. 
To this end, we choose  kinematics of an irreducible background to Higgs boson production 
and take  the center-of-mass energy $\sqrt{s}$ to be  the 
mass of the Higgs boson $\sqrt{s} = m_H = 125~{\rm GeV}$. The invariant mass of the vector boson $V_1$ is set to 
$p_3^2 = m_W^2$, with $m_W =  80.419~{\rm GeV}$. The invariant mass of the second vector boson $V_2$ is set to $25~{\rm GeV}$.
We take the vector boson scattering angle in the center-of-mass collision frame to be $\pi/3$ radians. 
We also take decay angles of the lepton $l_5$ in the rest frame of the boson $V_1$ to be $\theta_5 = \pi/4$ 
and $\varphi_5 = \pi/2$ and decay angles of the lepton $l_7$ in the rest frame of the boson $V_2$ 
to be $\theta_7 = \pi/6$ and $\varphi_7 = \pi$.  The four-momenta of initial and final state 
particles are given by 
\be
\label{eq22}
\begin{split}
& p_1 = (  62.5,0,0,62.5),\;\;\;\;\;  p_2 = (62.5,0,0,-62.5),\\
& p_5 = ( 
   48.2561024468725,\;        13.8697156788798,\;       -28.4324101181205,\;     
   36.4400941989053  ), \\ 
& p_6 = ( 
   37.6127597971275,\;        12.2010429705974,\;        28.4324101181205,\;     
  -21.3881346746519     
), \\     
& p_7 = ( 
   19.5655688780000 ,\;      -19.2853793247386,\;     0,\;
   3.29933778517879     
),\\     
& p_8 = (
  19.5655688780000,\;       -6.78537932473856, \;      0, \;
  -18.3512973094322 
).  
\end{split} 
\ee

Our results for leading and next-to-leading order helicity amplitudes at the kinematic point Eq.~(\ref{eq22}) 
are shown in Table~\ref{table:twoloopampl}. The divergent terms of the next-to-leading 
order amplitude are  compared to predictions based on Eq.(\ref{eqdiv}) and perfect 
agreement is found.  We also compared the  leading order helicity amplitude with the  results of previous 
computations~\cite{glover,matsuura,zecher,Binoth:2005ua}, as implemented in the 
program {\sf MCFM}~\cite{mcfm},  and found agreement.  Finally, we note that we compared the numerical 
results for helicity amplitudes reported in Table~\ref{table:twoloopampl} with the results 
of the independent calculation \cite{lorenzo1} and found complete agreement. 

\begin{table}[t]
\begin{center}
\begin{tabular}{|l|c|c|c|c|}
\hline
Helicity & ${\cal A}_1(\ep = 0)$  & ${\cal A}^{(2)}/{\cal A}_1(\ep = 0)$, $\; 1/\ep^2 \;$ & 
${\cal A}^{(2)}/{\cal A}_1(\ep = 0)$, $1/\ep$ & ${\cal A}^{(2)}/{\cal A}_1(\ep = 0)$, $\ep^0$  \\
\hline
LLLL &  
$-5169.9932 + i\; 10017.414  $ &
$-3.0 $  &
$-9.45694415 - i\; 16.4895884  $  & 
$ 42.4852911 - i\; 65.01495  $
 \\
\hline
LRLL &  
$ -6427.41534 - i\; 2610.6160  $ &
$-3.0 $  &
$-14.9422361 - i\; 9.2198662   $ & 
$-6.774561932 - i\; 71.66763   $
 \\
\hline
\end{tabular}
\end{center}
\caption{
Leading and next-to-leading order helicity amplitudes. Momenta of external particles are given in the main text of the paper. 
 }
\label{table:twoloopampl}
\end{table}

\begin{figure}[t]
  \centering
  \includegraphics[angle=-90,width=0.6\textwidth]{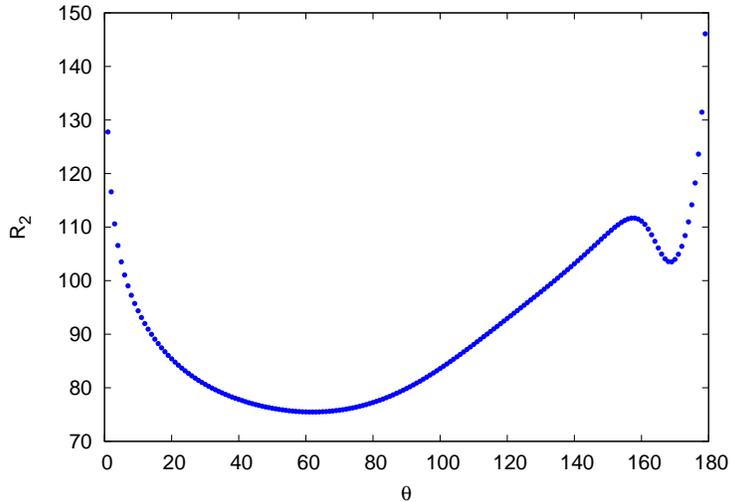}
  \caption{The ratio of finite parts of two- and one-loop helicity amplitudes 
$ R_2 = | A_2(LL) + A_2(LR)|/(|A_1(LL)| + |A_1(LR)|) $  as a function  of the vector-boson scattering angle. }
  \label{fig2}
\end{figure}

Finally, it is interesting to explore the numerical stability of $gg \to V_1 V_2$  amplitudes that we computed in this paper. 
The numerical stability  of such amplitudes is  known to be a potentially sensitive issue as earlier one-loop studies showed, 
see e.g. Ref.~\cite{Campbell:2013una, Binoth:2005ua}.  To study numerical stability, 
we consider the same kinematic point as described above but we treat  the vector boson 
scattering angle as a free parameter. We then compare the results of the double-precision 
implementation of the form factors in a {\sf Fortran} program with, effectively, 
arbitrary-precision calculation in {\sf Mathematica}. We find that 
 helicity amplitudes computed in these two different ways agree well 
for values of the scattering angle as small (large) as  $\theta = 1$ ( $\theta = 179$)  degrees. 
For such angles, the transverse momentum of a vector boson is just $0.5~{\rm GeV}$. 
We therefore conclude that our numerical implementation of  helicity amplitudes 
is sufficiently stable to allow their use in realistic numerical calculations. 
As a further illustration of these numerical results, in Fig.~\ref{fig2} we show absolute values of helicity
amplitudes as a function of the scattering angle. 

\section{Conclusions} 
\label{concl}

In this paper, we computed the helicity amplitudes for the production of  electroweak gauge bosons in gluon fusion $gg \to V_1 V_2$. 
The electroweak gauge bosons are allowed to have different masses and be off-shell; their decays to fermion pairs 
are taken into account explicitly.  The helicity amplitudes for $gg \to V_1 V_2$ are described by nine helicity-dependent 
form factors. We construct projection operators to compute those form factors from Feynman diagrams,
using integration-by-parts identities and the master integrals calculated  by us previously. 
Analytic results for helicity amplitudes are implemented in a {\sf Fortran} code that is available 
from the authors upon request. 

The results for the scattering amplitudes $gg  \to W^+ W^- $   and $gg  \to ZZ  $  
obtained in this paper open up an opportunity to compute the  NLO QCD corrections 
to the production of a pair of electroweak vector boson  in gluon fusion. Such calculations 
are interesting both at low and high center-of-mass collision energies. In the former case, 
they allow for a more accurate estimate of the irreducible background to Higgs production 
and the possible signal/background interference effects. In the latter case, precise 
predictions for $gg \to ZZ$ are important for improving prospects of constraining total Higgs 
boson decay width following Refs.~\cite{Caola:2013yja,Khachatryan:2014iha,atlas}, as well as for more generic
Higgs off-shell studies~\cite{Kauer:2012hd}. The corresponding 
real emission corrections  $gg \to ZZg$ have to be calculated in this case.  Such calculations 
are clearly possible with the existing one-loop technology~\cite{Campbell:2014gua,Melia:2012zg,agr,paco} 
and will be done in the near future.

\section*{Acknowledgment} 
K.M. is grateful to L.~Tancredi for useful conversations. F.C. would like to thank 
the  Institute for Advanced Studies, Princeton, 
for the hospitality  extended to him during completion of this paper.  
We are grateful to  A.~von~Manteuffel and L.~Tancredi for the opportunity  to 
compare results for $gg \to V_1 V_2$ amplitude   prior to  their publication. 
J.M.H. is supported in part by the DOE grant DE-SC0009988 and by 
the Marvin L. Goldberger fund.  
The research of K.M. was supported by Karlsruhe Institute of Technology through its startup grant. 
A.S. is  supported in part by DFG through SFB/TR 9.
V.S. is  supported in part by the Alexander von Humboldt Foundation (Humboldt Forschungspreis).

\appendix 

\section{The relations between various form factors} 

For the two possible helicity combinations, the invariant form factors, expressed through $T$ form-factors, read
\be
\begin{split}
F^{LL}_1 &= \frac{1}{2} \left[ \left( m_3^2 - u \right) (T_{12}+T_{8})-\left( m_4^2 - t \right) (T_{10}+T_{14})-s (p_\perp^2 T_{17}+2 T_{4})\right],
\\
F^{LL}_2 &= \frac{1}{2} \left[\left( m_3^2 
- t \right) (T_{12}+T_{8})-\left( m_4^2 - t \right) (T_{11}+T_{15})-p_\perp^2 s T_{19}-2 s T_{6}+2 T_{2}+2 T_{3}\right],
\\
F^{LL}_3 &= \frac{1}{2} \left[\left( m_3^2 - u \right) (T_{13}+T_{9})-\left( m_4^2 - u \right) (T_{10}+T_{14})-p_\perp^2 s T_{18}-2 s T_{5}+2 T_{2}+2 T_{3}\right],
\\
F^{LL}_4 &= \frac{1}{2} \left[\left( m_3^2 - t \right) (T_{13}+T_{9})-\left( m_4^2 - u \right) (T_{11}+T_{15})-s (p_\perp^2 T_{20}+2 T_{7})\right],
\\
F^{LL}_5 &= -\frac{1}{2} s \left(p_\perp^2 T_{16}+2 T_{1}+T_{2}+T_{3}\right),
~~
F^{LL}_6 = \frac{\left( m_3^2 - u \right) (T_{3}-T_{2})}{p_\perp^2 s}-T_{10}+T_{14},
\\
F^{LL}_7 &= \frac{\left( m_3^2 - t \right) (T_{3}-T_{2})}{p_\perp^2 s}-T_{11}+T_{15},
~~
F^{LL}_8 = \frac{\left( m_4^2 - t \right) (T_{3}-T_{2})}{p_\perp^2 s}+T_{12}-T_{8},
\\
F^{LL}_9 &= \frac{\left( m_4^2 - u \right) (T_{3}-T_{2})}{p_\perp^2 s}+T_{13}-T_{9};
\\
F^{LR}_1 &= \frac{1}{2} \left[\left(m_4^2-t\right) \left({p_\perp^2} s (T_{10}+T_{14})
-2 \left(m_3^2-u\right) (T_{2}+T_{3})\right)+{p_\perp^2} s \left({p_\perp^2} s T_{17}-\left(m_3^2-u\right) (T_{12}+T_{8})\right)\right],
\\
F^{LR}_2 &= \frac{1}{2} \left[ \left( m_4^2-t\right) \left( {p_\perp^2} s (T_{11}+T_{15})
-2 \left(m_3^2-t\right) (T_{2}+T_{3})\right)-{p_\perp^2} s \left( \left(m_3^2-t\right) (T_{12}+T_{8})
\right. \right.
\\
& \left. \left.  -{p_\perp^2} s T_{19}-2 (T_{2}+T_{3})\right)\right],
\\
F^{LR}_3 &= \frac{1}{2}{ p_\perp^2} s \left( -\left(m_3^2-u\right) (T_{13}+T_{9})+
\left(m_4^2-u\right) (T_{10}+T_{14})+{p_\perp^2} s T_{18}\right)
\\
&   +(T_{2}+T_{3}) \left( {p_\perp^2} s-\left(m_3^2-u\right) \left(m_4^2-u\right)\right)  ,
\\
F^{LR}_4 &= \frac{1}{2} (\left(m_4^2-u\right) ({p_\perp^2} s (T_{11}+T_{15})
-2 \left(m_3^2-t\right) (T_{2}+T_{3}))+{p_\perp^2} s ({p_\perp^2} s T_{20}-\left(m_3^2-t\right) (T_{13}+T_{9}))),
\\
F^{LR}_5 &= \frac{1}{2} {p_\perp^2} s^2 ({p_\perp^2} T_{16}-T_{2}-T_{3}),
~~
F^{LR}_6 = {p_\perp^2} s (T_{10}+T_{14})-\left(m_3^2-u\right) (T_{2}+T_{3}),
\\
F^{LR}_7 &= {p_\perp^2} s (T_{11}+T_{15})-\left(m_3^2-t\right) (T_{2}+T_{3}),
~~
F^{LR}_8 = \left(m_4^2-t\right) (T_{2}+T_{3})+{p_\perp^2} s (T_{12}+T_{8}),
\\
F^{LR}_9 &= \left(m_4^2-u\right) (T_{2}+T_{3})+{p_\perp^2} s (T_{13}+T_{9}).
\end{split} 
\ee

The square of the transverse momentum written through Mandelstam invariants reads  $p_\perp^2 = -(tu - m_3^2 m_4^2)/s$.

When the $F$ form factors are expressed in terms of projections $G_{1,..,20}$ shown earlier, the results appear to be relatively simple.
In particular, all spurious poles in $d-4$, present in the relations between $T$'s and $G$'s  cancel out in 
the relations between $F$'s and $G$'s. To give an example of these relations, we show results 
for a few  of the simplest form factors for the $LL$ amplitude
\be
\begin{split}
\\
& 
F_6^{LL} =  \frac{(m_3^2 - u)(G_{12}+G_{16})}{(d-3)(m_3^2-t)} + \frac{s (m_3^2+p_\perp^2)( G_{13} + G_{17} )}{(d-3)(m_3^2-t)},
\\
& F_7^{LL} = \frac{G_{12} + G_{16}  + (t-m_{3}^2)( G_{13} + G_{17} )}{d-3},
\\  
& F_{8}^{LL} = \frac{ (m_4^2-t) G_{10} -s(m_4^2+p_\perp^2) G_{11} }{(d-3)(m_4^2-u)}
+ \frac{G_{14} + (m_4^2 - t) G_{15}}{d-3},
\\
&
F_{9}^{LL} = \frac{(u-m_{4}^2) G_{11} - G_{10} }{d-3} + \frac{(u-m_4^2)G_{14} + s (m_4^2+p_\perp^2) G_{15}  }{(d-3)(m_4^2-t)}.
\end{split}
\ee

\end{document}